\documentclass[showpacs,preprintnumbers,amsmath,amssymb]{revtex4}
\usepackage{graphicx}
\usepackage{dcolumn}
\usepackage{bm}
\newcommand{\omu}{\overline{\mu}}
\newcommand{\e}{\epsilon}
\newcommand{\dl}{\delta}

\newcommand{\og}{\overline{g}}

\newcommand{\p}{\partial}
\newcommand{\f}{\frac}
\begin{document}
\preprint{RUG preprint}
\title{On ghost condensation, mass generation and Abelian dominance in the Maximal Abelian Gauge}
\author{David Dudal}
    \altaffiliation{Research Assistant of the Fund For Scientific
Research-Flanders (Belgium)}
    \email{david.dudal@rug.ac.be}
\author{Henri Verschelde}
 \email{henri.verschelde@rug.ac.be}
\affiliation{Ghent University
\\ Department of Mathematical
Physics and Astronomy \\ Krijgslaan 281-S9 \\ B-9000 GENT,
BELGIUM}
\begin{abstract}
Recent work claimed that the off-diagonal gluons (and ghosts) in
pure Yang-Mills theories, with Maximal Abelian gauge fixing (MAG),
attain a dynamical mass through an off-diagonal ghost condensate.
This condensation takes place due to a quartic ghost interaction,
unavoidably present in MAG for renormalizability purposes. The
off-diagonal mass can be seen as evidence for Abelian dominance.
We discuss why ghost condensation of the type discussed in those
works cannot be the reason for the off-diagonal mass and Abelian
dominance, since it results in a tachyonic mass. We also point out
what the full mechanism behind the generation of a real mass might
look like.
\end{abstract}
 \maketitle
\section{\label{sec1}Introduction}
As everybody knows, quarks are confined: nature as well as lattice
simulations of nature are telling us that. Still, there is no
rigorous proof of confinement. One proposal for the explanation of
confinement is the idea of the dual superconductor: magnetic
monopoles condense and induce a dual Meissner effect:
color-electric flux between charges is squeezed and a string is
created in between. The original work on this topic can be found
in \cite{Nambu:1974zg,Mandelstam:1974pi,thooft}. Abelian
projection \cite{'tHooft:1981ht} is a way to reveal the relevant
degrees of freedom (the monopoles). In a lose way of speaking, at
points were the projection is ill-defined, singularities invoke
(Abelian) monopoles. Abelian dominance means that low energy QCD
is dominated by Abelian degrees of freedom. Some early work on
this is presented in \cite{Ezawa:bf}. Numerical evidence can be
found in e.g. \cite{Suzuki:1992gz,Hioki:1991ai,Suzuki:1989gp} and
more recently \cite{Amemiya:zf}. \\\\Can this Abelian dominance be
founded on more theoretical grounds? In the light of
renormalization $\grave{\textrm{a}}$ la Wilson, and assuming that
the \emph{off-diagonal} gluons (ghosts) attain a mass $M$ while
the \emph{diagonal} ones remain massless, an effective theory in
terms of the massless diagonal fields could be achieved at low
energy ($\ll M$), thereby realizing a kind of Abelian dominance.
In the context of low energy theories, we like to refer to the
Appelquist-Carazzone decoupling theorem \cite{Appelquist:tg},
which states that heavy particle modes decouple at low energy.
Notice that this decoupling does not mean "heavy terms" are simply
removed by hand from the Lagrangian, their influence is still
present through renormalization effects. As an illustration of
this: a low energy, \emph{Abelian} theory for Yang-Mills was
derived in \cite{Kondo:2000sh}, but the corresponding
$\beta$-function was shown
to be the same as the full Yang-Mills one.\\\\
The aforementioned pathway has been followed in a series of papers
by Kondo et al
\cite{Kondo:2000sh,Kondo:2000zv,Kondo:2000ey,Kondo:1999tj,Kondo:1999je,Kondo:1997kn,Kondo:1998sr,
Kondo:1997pc} and more recently the technique of the exact
renormalization group has been employed by Freire
\cite{Freire:2002ba,Freire:2001nd} to construct effective low
energy descriptions of Yang-Mills theory. The results have been
used in order to construct a linearly rising potential between
static quarks, a criterion for confinement. Their efforts were
based on the dual superconductor picture, realized with MAG. Also
the monopole condensation was discussed in their framework. An
essential ingredient of their work is the mass scale of the
off-diagonal fields. The monopole condensate is proportional to
this mass squared \cite{Kondo:2000sh}. The lattice reported a
value of approximately 1.2 GeV for the off-diagonal gluon mass in
MAG Yang-Mills \cite{Amemiya:zf}. Next to these numerical results,
analytical information is needed how this mass raises. A few
papers have been written on this issue
\cite{Kondo:2000ey,Schaden:2001xu,Schaden:2000fv,Schaden:1999ew}.
All these authors came to the same conclusion: $\emph{a dimension
two ghost condensation gives an off-diagonal mass M}$. We already
mentioned (but did not show explicitly) in a previous paper that
we found the ghost condensation gives a \emph{tachyonic}
off-diagonal gluon mass \cite{Dudal:2002aj}. In this paper, we
will perform the calculations explicitly step by step. To make it
self-contained, we will start from the beginning and in order to
make comparison as transparent as possible, we will follow the
(notational) conventions of \cite{Kondo:2000ey}. For the sake of
simplicity, we will restrict ourselves to the SU(2) case. We
discuss the (in)completeness of presented work. We end with the
path we intend to follow in the future to investigate dynamical
mass generation in MAG.

\section{\label{sec2}Ghost condensation in the Maximal Abelian Gauge}
Consider the Yang-Mills Lagrangian in four-dimensional Minkowski
space time
\begin{equation}\label{1}
    \mathcal{L}=-\f{1}{4}\mathcal{F}_{\mu\nu}^{A}\mathcal{F}^{A\mu \nu}+\mathcal{L}_{GF+FP}
\end{equation}
where $\mathcal{L}_{GF+FP}$ is the gauge fixing and Faddeev-Popov
part. \\We decompose the gauge field as
\begin{eqnarray}
    \label{2a}\mathcal{A}_{\mu}&=&\mathcal{A}_{\mu}^{A}T^{A}=a_{\mu}T^{3}+A_{\mu}^{a}T^{a}\\
    \label{2b}\mathcal{F}_{\mu\nu}&=&\mathcal{F}_{\mu\nu}^{A}T^{A}=\mathcal{D}_{\mu}\mathcal{A}_{\nu}-\mathcal{D}_{\nu}\mathcal{A}_{\mu}=\p_{\mu}\mathcal{A}_{\nu}-\p_{\nu}\mathcal{A}_{\mu}-ig\left[\mathcal{A}_{\mu},\mathcal{A}_{\nu}\right]
\end{eqnarray}
The $T^{A}$'s are the Hermitian generators of SU(2) and obey the
commutation relations $\left[T^{A},T^{B}\right]=if^{ABC}T^{C}$.
$T^{3}$ is the diagonal generator. The capital index $A$ runs from
1 to 3. Small indices like $a$, $b$,... run from 1 to 2 and label
the off-diagonal components. We will drop the index 3 later on.
\\As a gauge fixing procedure, we use MAG. Introducing the
functional
\begin{equation}\label{3}
    \mathcal{R}[A]=(VT)^{-1}\int d^{4}x\left(\f{1}{2}A_{\mu}^{a}A^{\mu a}\right)
\end{equation}
with $VT$ the space time volume, MAG is defined as that gauge
which minimizes $\mathcal{R}$ under local gauge transformations.
Since (\ref{3}) is invariant under U(1) transformations w.r.t. the
"photon" $a_{\mu}$, MAG is only a partial gauge fixing. We do not
fix the residual U(1) gauge freedom, since it plays no role for
what we are discussing here.\\ To implement the gauge fixing in
the Lagrangian (\ref{1}), we use the so-called modified MAG. This
gauge is slightly different from the ordinary MAG, it possesses
for instance some more symmetry (see \cite{Kondo:1998sr} and
references therein). Moreover, it generates the four-point ghost
interaction, indispensable for the renormalizibility of MAG, as
was proven in \cite{Min:bx}.\\Explicitly, we get
\begin{equation}\label{6}
    \mathcal{L}_{GF+FP}=i\dl_{B}\overline{\dl}_{B}\left(\f{1}{2}A_{\mu}^{a}A^{\mu a}-\f{\alpha}{2}iC^{a}\overline{C}^{a}\right)
\end{equation}
where $\alpha$ is a gauge parameter, $C$ and $\overline{C}$ denote
the (off-diagonal) ghosts and anti-ghosts, $\dl_{B}$ and
$\overline{\dl}_{B}$ are the BRST and anti-BRST transformation
respectively, defined by \footnote{$\mathcal{C}=(C^{a},C^{3})$
with $C^{3}$ the diagonal ghost. Analogously for
$\overline{\mathcal{C}}$.}
\begin{eqnarray}\label{7}
    \dl_{B}\mathcal{A}_{\mu}&=&\mathcal{D}_{\mu}\mathcal{C}=\p_{\mu}\mathcal{C}-ig\left[\mathcal{A}_{\mu},\mathcal{C}\right]\nonumber\\
    \dl_{B}\mathcal{C}&=&\f{ig}{2}\left[\mathcal{C},\mathcal{C}\right]\nonumber\\
    \dl_{B}\overline{\mathcal{C}}&=&i\mathcal{B}\nonumber\\
    \dl_{B}\mathcal{B}&=&0\\\nonumber\\
    \overline{\dl}_{B}\mathcal{A}_{\mu}&=&\mathcal{D}_{\mu}\overline{\mathcal{C}}=\p_{\mu}\overline{\mathcal{C}}-ig\left[\mathcal{A}_{\mu},\overline{\mathcal{C}}\right]\nonumber\\
    \overline{\dl}_{B}\overline{\mathcal{C}}&=&\f{ig}{2}\left[\overline{\mathcal{C}},\overline{\mathcal{C}}\right]\nonumber\\
    \overline{\dl}_{B}\mathcal{C}&=&i\overline{\mathcal{B}}\nonumber\\
    \overline{\dl}_{B}\overline{\mathcal{B}}&=&0\nonumber\\
    \mathcal{B}+\overline{\mathcal{B}}&=&g\left[\mathcal{C},\overline{\mathcal{C}}\right]
\end{eqnarray}
with the following properties
\begin{eqnarray}\label{8}
    \dl_{B}^{2}&=&\overline{\dl}_{B}^{2}=\left\{\dl_{B},\overline{\dl}_{B}\right\}=0\nonumber\\
    \dl_{B}\left(XY\right)&=&\dl_{B}\left(X\right)Y\pm
    X\dl_{B}\left(Y\right)\nonumber\\
    \overline{\dl}_{B}\left(XY\right)&=&\overline{\dl}_{B}\left(X\right)Y\pm
    X\overline{\dl}_{B}\left(Y\right)
\end{eqnarray}
where the upper sign is taken for bosonic $X$, and the lower sign
for fermionic $X$.\\\\ Performing the BRST and anti-BRST
transformations, yields
\begin{eqnarray}\label{8bis}
    \mathcal{L}_{GF+FP}&=&B^{a}D_{\mu}^{ab}A^{\mu
    b}+\f{\alpha}{2}B^{a}B^{a}+i\overline{C}^{a}D_{\mu}^{ac}D^{\mu
    cb}C^{b}-ig^{2}\e^{ad}\e^{cb}\overline{C}^{a}C^{b}A^{\mu c}A_
    {\mu}^{d}\nonumber\\&+&i\overline{C}^{a}g\e^{ab}C^{3}D_{\mu}^{bc}A^{\mu
    c}-i\alpha
    g\e^{ab}B^{a}\overline{C}^{b}C^{3}+\f{\alpha}{4}g^{2}\e^{ab}\e^{cd}\overline{C}^{a}\overline{C}^{b}C^{c}C^{d}
\end{eqnarray}
where
\begin{equation}\label{9tris}
    D_{\mu}^{ab}\equiv
    D_{\mu}^{ab}[a]=\p_{\mu}\dl^{ab}-g\e^{ab}a_{\mu}
\end{equation}
is the covariant derivative w.r.t. the U(1) symmetry and
\begin{eqnarray}\label{9tris}
    \e^{12}&=&-\e^{21}=1\\
    \e^{11}&=&\e^{22}=0
\end{eqnarray}
When we integrate the multipliers $\mathcal{B}$ out, we finally
obtain
\begin{equation}\label{9}
    \mathcal{L}_{GF+FP}=-\f{1}{2\alpha}\left(D_{\mu}^{ab}A^{\mu
    b}\right)^{2}+i\overline{C}^{a}D_{\mu}^{ac}D^{\mu
    cb}C^{b}-ig^{2}\e^{ad}\e^{cb}\overline{C}^{a}C^{b}A^{\mu
    c}A_{\mu}^{d}+\f{\alpha}{4}g^{2}\e^{ab}\e^{cd}\overline{C}^{a}\overline{C}^{b}C^{c}C^{d}
\end{equation}
Notice that the diagonal ghost $C^{3}$ has dropped out of
(\ref{9}).\\For the (singular) choice $\alpha=0$, the 4-ghost
interaction cancels from the Lagrangian. However, radiative
corrections due to the other, non-vanishing 4-point interactions,
reintroduce this term. We further assume that $\alpha\neq0$. Some
more details concerning the properties for $\alpha=0$ can be found
in \cite{Schaden:1999ew}.
\\\\
To discuss the ghost condensation mechanism, we "Gaussianize" the
4-ghost interaction in the Lagrangian by means of the (U(1)
invariant) auxiliary field $\phi$
\begin{equation}\label{10}
    \f{\alpha}{4}g^{2}\e^{ab}\e^{cd}\overline{C}^{a}\overline{C}^{b}C^{c}C^{d}\rightarrow-\f{1}{2\alpha
    g^{2}}\phi^{2}-i\phi\e^{ab}\overline{C}^{a}C^{b}
\end{equation}
A useful identity to prove (\ref{10}), reads
\begin{equation}\label{11}
    \e^{ab}\e^{cd}\overline{C}^{a}\overline{C}^{b}C^{c}C^{d}=2\left(i\e^{ab}\overline{C}^{a}C^{b}\right)^{2}
\end{equation}
The part of the Lagrangian which concerns us for the moment is
\begin{equation}\label{12}
    \hat{\mathcal{L}}=i\overline{C}^{a}\p_{\mu}\p^{\mu}C^{a}-\f{1}{2\alpha
    g^{2}}\phi^{2}-i\phi\e^{ab}\overline{C}^{a}C^{b}
\end{equation}
Assuming constant $\phi$, we use the Coleman-Weinberg construction
\cite{Coleman:jx} of the effective potential $V(\phi)$. This means
we are summing all 1-loop (off-diagonal) ghost bubbles with any
number of $\phi$-insertions. This yields
\begin{equation}\label{13}
    (VT)V(\phi)=\int d^{4}x\f{\phi^{2}}{2\alpha
    g^{2}}+i\ln\det\left(\p_{\mu}\p^{\mu}\dl^{ab}-\phi\epsilon^{ab}\right)
\end{equation}
or
\begin{equation}\label{15}
    V(\phi)=\f{\phi^{2}}{2\alpha
    g^{2}}-\f{1}{i}\int \f{d^{4}k}{\left(2\pi\right)^{4}}\ln\left(
    k^{4}+\phi^{2}\right)
\end{equation}
Employing the Wick rotation $k_{0}\rightarrow ik_{0}$ \footnote{If
one would like to avoid Wick rotations, one could start
immediately from the Euclidean version of Yang-Mills.}, and
performing the integration in dimensional regularization within
the $\overline{MS}$ scheme, we arrive at
\begin{equation}\label{16}
    V(\phi)=\f{\phi^{2}}{2\alpha
    \og^{2}}+\f{\phi^{2}}{32\pi^{2}}\left(\ln\f{\phi^{2}}{\omu^{4}}-3\right)
\end{equation}
This potential possesses a local maximum at $\phi=0$ (the usual
vacuum), but has global minima at
\begin{equation}\label{17}
    \phi=\pm v=\pm \omu^{2}e^{1-\f{8\pi^{2}}{\alpha \og^{2}(\omu^{2})}}
\end{equation}
We take $\alpha>0$ since $v$ diverges for $\og^{2}\rightarrow0$ if
$\alpha<0$. \\\\Up to now, we find complete agreement with
\cite{Kondo:2000ey}. We proceed by calculating the ghost
propagator in the non-zero vacuum ($V(v)<0$). Substituting
$\phi=v$ in (\ref{12}), it is straightforward to determine the
Feynman propagator
\begin{equation}\label{18}
    \left\langle C^{a}(x)\overline{C}^{b}(y)\right\rangle=\int
    \f{d^{4}k}{\left(2\pi\right)^{4}}\f{-k^{2}\dl^{ab}+v\e^{ab}}{k^{4}+v^{2}}e^{-ik(x-y)}
\end{equation}
With the above propagator, we are ready to determine the 1-loop
off-diagonal gauge boson polarization. Now, there exists a
non-trivial contribution coming from the ghost bubble, originating
in the interaction term
$-ig^{2}\e^{ad}\e^{cb}\overline{C}^{a}C^{b}A^{\mu
    c}A_{\mu}^{d}$, resulting in a mass $M$ for the off-diagonal
    gluons. Again
Wick rotating $k_{0}\rightarrow ik_{0}$ to get an integral over
Euclidean space time, one easily obtains
\begin{equation}\label{dudal1}
    M^2=g^2\int
    \f{d^{4}k}{\left(2\pi\right)^{4}}\f{2k^{2}}{k^{4}+v^{2}}
\end{equation}
There is one remaining step, we still have to calculate the
integral of (\ref{dudal1}). Using dimensional regularization, we
find the finite result
\begin{equation}\label{23}
    M^2=\f{-g^2v}{16\pi}<0
\end{equation}
where we have used that $v>0$. Here we find a \emph{different}
result in comparison with the other references
\cite{Kondo:2000ey,Schaden:2001xu,Schaden:2000fv,Schaden:1999ew}.
To be more precise, we find the opposite sign. This sign
difference is not meaningless, since the negative sign we find
means that the off-diagonal fields have a \emph{tachyonic} mass.
\\\\
Hence, we state that a ghost condensation $\grave{\textrm{a}}$ la
$\left\langle\epsilon^{ab}\overline{C}^{a}C^{b}\right\rangle$ is
\emph{not} the mechanism behind the off-diagonal mass generation
in MAG, and consequently does not give evidence for Abelian
dominance.\\
\\Another important point is what happens with the diagonal gluon. Consider the
term $i\overline{C}^{a}D_{\mu}^{ac}D^{\mu cb}C^{b}$ of (\ref{9}),
it contains a part proportional to $i\overline{C}^{a}C^{a}
a_{\mu}a^{\mu}$. Doing the same as for the off-diagonal gluons,
the diagonal gluon $a_{\mu}$ seems to get a (real) mass too, which
is of the same order as the off-diagonal one (up to the sign).
However, there are other 1-loop contributions coming from the
terms proportional to
$\e^{ab}\left(\p_{\mu}\overline{C}^{a}\right)C^{b}a^{\mu}$ and
$\e^{ab}\overline{C}^{a}\left(\p_{\mu}C^{b}\right)a^{\mu}$. These
contributions cancel the one coming from the term proportional to
$i\overline{C}^{a}C^{a} a_{\mu}a^{\mu}$. Consequently, the
"photon" $a_{\mu}$ remains massless, as could be expected by the
residual U(1) invariance.\\\\ Another point of concern is the
renormalizibility of the "Gaussianized" Lagrangian. A completely
analogous approach can be done in case of the 2-dimensional
Gross-Neveu model \cite{Gross:jv}, where the 4-fermion interaction
can also be made Gaussian by the introduction of an auxiliary
field $\sigma$. This works well at 1-loop order, but from 2 loops
on, \emph{ad hoc} counterterms have to be added in order to end up
with finite results \cite{Luperini:1991sv}. A successful formalism
to deal with local composite operators in case of the Gross-Neveu
model was developed in \cite{Verschelde:jx}. A similar approach
should be used to investigate the ghost condensates. \\\\One could
wonder what the mechanism behind the mass generation might be,
since the previous paragraphs showed that we didn't find a
dynamically generated real mass for the (off-diagonal) particles.
In order to find an answer to this question, we first give a very
short overview of recent results in the Landau gauge, giving us a
hint in which direction we should look for the mass generation.

\section{\label{sec3} Gluon condensation via $\mathcal{A}^{2}$ in the Landau gauge and its nephew $A^{2}$ in the Maximal Abelian
Gauge}

A well known condensate in QCD (or Yang-Mills) is the dimension
four gluon condensate
$\left\langle\mathcal{F_{\mu\nu}^{A}}\mathcal{F^{\mu\nu
A}}\right\rangle$. This is the lowest dimensional gluonic
condensate that can exist, since no local, gauge-invariant
condensates with dimension lower than 4 exist. However, recently
interest arised concerning a dimension 2 gluon condensate in
Yang-Mills theory in the Landau gauge. One way it came to
attention was the conclusion that there exists a non-negligible
discrepancy between the lattice strong coupling constant
$\alpha_{s}$ (determined via the 3-point gluon interaction) and
the perturbative one, into a relatively high energy region where
this wouldn't be expected (up to 10 GeV). Also the propagator
showed a similar discrepancy. The $\f{1}{p^{4}}$ power correction
due to $\left\langle \mathcal{F}^{2}\right\rangle$ is far to small
to explain this. It was shown that a $\f{1}{p^{2}}$ power
correction could solve the discrepancy. More precisely, the
Operator Product Expansion (OPE) used in combination with the
$\left\langle \mathcal{A}^{2}\right\rangle$ condensate was able to
fit both predictions \cite{Boucaud:2000,Boucaud:2001st}. An
important question that naturally arises, sounds: has
$\left\langle \mathcal{A}^{2}\right\rangle$ any physical meaning,
or is it merely a gauge artefact? The point is that
$\mathcal{A}^{2}$ equals $(VT)^{-1}\min_{U}\int d^{4}x
\mathcal{A}_{\mu}^{U}\mathcal{A}^{\mu U}$ in the Landau gauge, and
this latter operator is, although non-local, gauge-invariant.
Hence, $\mathcal{A}^{2}$ can be given some physical sense in the
Landau gauge. Moreover, \cite{Gubarev:2000eu} discussed the
relevance of $\mathcal{A}^{2}$ in connection with topological
structure (monopoles) of compact QED. The physical relevance of
the Landau gauge, in the framework of geometrical monopoles, is
explained in \cite{Gubarev:2000nz}. The authors of that paper also
stress that the values found with an OPE calculation, only
describe the soft (infrared) content of $\left\langle
\mathcal{A}^{2}\right\rangle$, while they argue that also hard
(short range) contributions, unaccessible for OPE, may occur. In
this context, we cite \cite{Verschelde:2001ia}, where a formalism
was constructed for the calculation of the vacuum expectation
value of (local) \emph{composite} operators. Since this is based
on the effective action, it should in principle, give the "full"
value of $\left\langle \mathcal{A}^{2}\right\rangle$, i.e. soft
and hard part. For example, one could assume an instanton
background as a possible source of long range contributions. In
fact, there is some preliminary evidence that instantons can
explain the OPE values \cite{Boucaud:2002nc}.\\The conclusion that
one can draw from all this is that the dimension 2 condensate
$\left\langle\mathcal{A}^{2}\right\rangle$ may have some physical
relevance in the Landau gauge. \\\\Let us go back to MAG
\footnote{Here, with MAG we mean the gauge minimizing the
functional (\ref{3}), and not the modified MAG.} now. In this
particular gauge, $(VT)^{-1}\min_{U}\int d^{4}x
\mathcal{A}_{\mu}^{U}\mathcal{A}^{\mu U}$ no longer reduces to a
local operator. It would be interesting to repeat e.g. the OPE
calculations of \cite{Boucaud:2001st} for the coupling constant
and propagators in MAG, but which dimension 2 condensate(s) could
take over the role of  $\left\langle \mathcal{A}^{2}\right\rangle$
in the Landau gauge? To solve this, we draw attention to the
striking similarity existing between the Landau gauge and MAG. The
former one can be seen as that gauge minimizing $(VT)^{-1}\int
d^{4}x \left(\mathcal{A}_{\mu}^{A}\mathcal{A}^{\mu A}\right)^{U}$,
while the latter one minimizes $(VT)^{-1}\int d^{4}x
\left(A_{\mu}^{a}A^{\mu a}\right)^{U}$. This operator reduces to
the local one $A^{2}$ in MAG and can be seen as the MAG version of
$\mathcal{A}^{2}$. Due to the more complex nature of the
(renormalizable) modified MAG, other dimension 2 condensates exist
(the ghost condensates). Notice that all these condensates are
U(1) invariants, hence the U(1) symmetry will be preserved.
\\\\The physics we see behind all these condensates is that they
might have a common, deeper reason for existence. In this context,
we quote \cite{Fukuda:1977mz,Fukuda:1977zp}, where it was shown
that the zero vacuum is instable (tachyonic) and a vacuum with
lower energy is achieved through gluon pairing, and an
accompanying gluon mass. The vacuum energy itself is a physical
object. After choosing a certain gauge, the different types of
dimension 2 condensates are just an expression of the fact that
$E=0$ is a wrong vacuum state. In this sense, all these different
condensates in different gauges are equivalent in a way, since
they lower the vacuum energy to a stable $E<0$ vacuum. In
\cite{Gubarev:2000nz}, discussion can be found on the appearance
of the soft part of $\left\langle \mathcal{A}^{2}\right\rangle$ in
gauge-variant quantities like in an OPE improvement of the gluon
propagator, while the hard part enters physical quantities. It is
imaginable that the mechanism behind this hard part (see
\cite{Gubarev:2000nz} for more details) is the same in different
gauges, but reveals its importance with different condensates,
depending on the specific gauge. This might justify its possible
appearance in gauge-invariant quantities.
\section{\label{sec4} Further discussion on the ghost condensation and mass generation in the modified MAG}
In \cite{Dudal:2002ye}, it was shown that it is possible to fix
the residual abelian gauge freedom of MAG in such a way that the
ghost condensate
$\left\langle\epsilon^{ab}\overline{C}^{a}C^{b}\right\rangle$ does
not give rise to any mass term. This abelian gauge fixing (needed
for a complete quantization of the theory) was based on the
requirement that the fully gauge fixed Lagrangian has a SL(2,R)
and anti-BRST invariance. A restricted \footnote{By restricted, we
mean that the symmetry only acts non-trivially on the off-diagonal
fields.}version of this SL(2,R) symmetry was originally observed
in SU(2) MAG in \cite{Schaden:1999ew}, and later generalized to
SU(N) MAG \cite{Sawayanagi:zw}. In \cite{Dudal:2002ye}, the
symmetry was defined on all the fields (diagonal and
off-diagonal). In fact, that SL(2,R) symmetry together with the
(anti-) BRST symmetry form a larger algebra, the Nakanishi-Ojima
(NO) algebra. This NO algebra is known to generate a symmetry of
the Landau gauge and a certain class of generalized covariant
gauges, more precisely the Curci-Ferrari gauges, given by the
gauge fixing Lagrangian
\begin{equation}\label{6CF}
    \mathcal{L}_{GF+FP}=i\dl_{B}\overline{\dl}_{B}\left(\f{1}{2}\mathcal{A}_{\mu}^{A}\mathcal{A}^{\mu A}-\f{\alpha}{2}i\mathcal{C}^{A}\overline{\mathcal{C}}^{A}\right)
\end{equation}
The Landau gauge corresponds to the gauge parameter choice
$\alpha=0$. For more details, see \cite{no,ni,oji,dj,ds}.
\\\\Yang-Mills theory with the gauge fixing (\ref{6CF}) possesses
a generalization to a massive SU(N) gauge model, the so-called
Curci-Ferrari model \cite{cf}. Although this model is non-unitary,
it is known to be (anti-)BRST invariant and renormalizible,
whereby the mass term is of the form
\begin{equation}\label{dudal10}
    \mathcal{L}_{mass}=M^{2}\left(\f{1}{2}\mathcal{A}_{\mu}^{A}\mathcal{A}^{\mu A}-i\alpha\mathcal{C}^{A}\overline{\mathcal{C}}^{A}\right)
\end{equation}
Keeping this in mind and recalling that in
\cite{Verschelde:2001ia}, a dynamically generated mass was found
in case of the Landau gauge by coupling a source $J$ to the
operator $A^{2}$, it becomes clear that in case of the
Curci-Ferrari gauge, the same technique could be employed by
coupling a source $J$ to the composite operator
\begin{equation}\label{dudal11}
    \mathcal{L}_{source}=J\left(\f{1}{2}\mathcal{A}_{\mu}^{A}\mathcal{A}^{\mu A}-i\alpha\mathcal{C}^{A}\overline{\mathcal{C}}^{A}\right)
\end{equation}
Returning to the case of MAG and comparing the gauge fixing
Lagrangians (\ref{6}) and (\ref{6CF}), the equivalent of
(\ref{dudal11}) reads
\begin{equation}\label{dudal12}
    \mathcal{L}_{source}=J\left(\f{1}{2}A_{\mu}^{a}A^{\mu a}-i\alpha C^{a}\overline{C}^{a}\right)
\end{equation}
This idea to arrive at a dynamically generated mass in case of the
Curci-Ferrari and Maximal Abelian gauge was already proposed in
\cite{Kondo:2001nq,Kondo:2001tm}. There, it was explicitly shown
that the operator coupled to the source $J$ in the expressions
(\ref{dudal11}) or (\ref{dudal12}), is on-shell BRST
invariant.\\We reserve the actual discussion of the aforementioned
framework to get a dynamical mass for future publications, since
it is quite involved and a clean treatment of it needs a
combination of the local composite operator formalism
\cite{Verschelde:2001ia} and the algebraic renormalization
technique \cite{book,bbh}.\\\\Before turning to conclusions, we
want to draw attention to the following. We decomposed the 4-ghost
interaction with a real auxiliary field $\phi$ whereby
$\phi\sim\epsilon^{ab}\overline{C}^{a}C^{b}$. Let's make a small
comparison with ordinary superconductivity. Usually, there is
talked about BCS pairing, i.e. particle-particle and hole-hole
pairing. The analogy of this in the ghost condensation case would
be ghost-ghost pairing and antighost-antighost pairing. This can
be achieved by an alternative decomposition of the 4-ghost
interaction via a pair of auxiliary fields $\sigma$ and
$\overline\sigma$ such that $\sigma\sim\epsilon^{ab}C^{a}C^{b}$
and
$\overline{\sigma}\sim\epsilon^{ab}\overline{C}^{a}\overline{C}^{b}$.
This kind of pairing \footnote{Our conclusion about the tachyonic
mass is unaltered by this alternative decomposition of the 4-ghost
interaction.} was considered in \cite{Lemes:2002ey}. A less known
effect is the particle-hole pairing, the so-called Overhauser
pairing \cite{bo}. This corresponds to the kind of condensation we
and the papers
\cite{Kondo:2000ey,Schaden:1999ew,Schaden:2000fv,Schaden:2001xu}
considered. From the viewpoint of the SL(2,R) symmetry, the
existence of different channels where the ghost condensation can
take place should not be suprising. The different composite ghost
operators are mutually changed into each other under the action of
the symmetry. Here and in the other papers the \emph{choice} was
made to work with the Overhauser channel, but a complete treatment
would need an analysis of all channels at once, and with the local
composite operator technique. This analysis of the BCS versus
Overhauser effect is nicely intertwined with the existence of the
NO algebra and its (partial) breakdown, and it is very much alike
for the MAG, Landau \cite{bezig} and Curci-Ferrari gauge, just as
in case of the mass generation mechanism. As an indication, it has
been found recently that, although no 4-ghost interaction is
present in the Landau gauge, the condensation $\grave{\textrm{a}}$
la $f^{ABC}\overline{\mathcal{C}}^{A}\mathcal{C}^{B}$ etc. also
occurs \cite{Lemes:2002rc}.

\section{\label{sec5} Conclusion}
We considered Yang-Mills theory in the Maximal Abelian Gauge. With
this non-linear gauge choice, a 4-ghost interaction enters the
Lagrangian. Such an interaction could allow a non-zero vacuum
expectation value for (off-diagonal) dimension 2 ghost
condensates. Consequently, it was expected that a mass generating
mechanism for the off-diagonal gluons \emph{and} the diagonal
gluons due to 4-point
interaction terms of the form gluon-gluon-ghost-anti-ghost was found. \\\\
We explained why this particular type of ghost condensation is not
sufficient to construct a (off-diagonal) dynamical mass in SU(2)
Yang-Mills theory in the Maximal Abelian Gauge, an indicator for
Abelian dominance. We have restricted ourselves to the SU(2) case,
but a similar conclusion will exist for general SU($N$). Explicit
calculations showed that we ended up with a tachyonic off-diagonal
mass $M$ ($M^{2}<0$). This result indicate something is missing. A
comparison with Yang-Mills theory in the Landau gauge and the role
played by the mass dimension 2 gluon condensate $\left\langle
\mathcal{A}^{2}\right\rangle$, shed some light on the route that
should be followed. \\\\ We revealed certain shortcomings of the
present available studies on the ghost condensation
(renormalizibility, existence of more than one condensation
channel).\\\\The actual study of the mass generation and the ghost
condensation with its symmetry breaking pattern will be discussed
elsewhere. We will follow the local composite operator formalism
of \cite{Verschelde:2001ia}, where a source is coupled to each
operator and the effective action can be treated consistently.
This effective potential formalism allows a clean treatment of the
role played by the dimension 2 operators. We remark that with
essentially perturbative techniques one can obtain at least
qualitatively trustworthy results \footnote{In case of the
Gross-Neveu model, very accurate results were obtained
\cite{Verschelde:jx}. In case of the $\left\langle
A^{2}\right\rangle$ condensate in the Landau gauge, the relevant
coupling constant was quite small, making the expansion acceptable
\cite{Verschelde:2001ia}.} on the stability of the condensates and
their relevance for e.g. mass generation and symmetry breakdown,
without making it directly necessary to go to (or extrapolating
to) strong coupling.
\\\\We conclude by mentioning that the dimension 2 condensates and
the accompanying mass generation in Yang-Mills are not only of
theoretical importance (the role of
$\left\langle\mathcal{A}^{2}\right\rangle$ for OPE corrections
\cite{Boucaud:2000,Boucaud:2001st}, monopoles
\cite{Gubarev:2000eu,Gubarev:2000nz}, \emph{short range linear
correction} to the Coulomb-like potential \cite{Boucaud:2000}, low
energy effective theories \cite{Kondo:2002xn},...) but also have
their importance for automated Feynmandiagram calculations
\cite{Gracey:2001,Browne:2002wd,Gracey:2002} where a gluon mass
serves as a infrared regulator. If this mass is generated in
massless Yang-Mills, it does not have to be implemented by hand.
\section{\label{sec1}Acknowledgments}
Private communication with M.~Schaden, S.~P.~Sorella and
J.~A.~Gracey is greatly acknowledged.


\begin{thebibliography}{99}
\bibitem{Nambu:1974zg}
Y.~Nambu, Phys.\ Rev.\ D {\bf 10} (1974) 4262

\bibitem{Mandelstam:1974pi}
S.~Mandelstam, Phys.\ Rept.\  {\bf 23} (1976) 245

\bibitem{thooft}
G.~'t Hooft, in: A.Zichichi (Ed.), High Energy Physics, Editorice
Compositori, Bologna (1975)

\bibitem{'tHooft:1981ht}
G.~'t Hooft, Nucl.\ Phys.\ B {\bf 190} (1981) 455

\bibitem{Ezawa:bf}
Z.~F.~Ezawa and A.~Iwazaki, Phys.\ Rev.\ D {\bf 25} (1982) 2681

\bibitem{Suzuki:1992gz}
T.~Suzuki, S.~Hioki, S.~Kitahara, S.~Kiura, Y.~Matsubara,
O.~Miyamura and S.~Ohno, Nucl.\ Phys.\ Proc.\ Suppl.\  {\bf 26}
(1992) 441

\bibitem{Hioki:1991ai}
S.~Hioki, S.~Kitahara, S.~Kiura, Y.~Matsubara, O.~Miyamura,
S.~Ohno and T.~Suzuki, Phys.\ Lett.\ B {\bf 272} (1991) 326
[Erratum-ibid.\ B {\bf 281} (1992) 416]

\bibitem{Suzuki:1989gp}
T.~Suzuki and I.~Yotsuyanagi, Phys.\ Rev.\ D {\bf 42} (1990) 4257

\bibitem{Amemiya:zf}
K.~Amemiya and H.~Suganuma, Phys.\ Rev.\ D {\bf 60} (1999) 114509

\bibitem{Appelquist:tg}
T.~Appelquist and J.~Carazzone, Phys.\ Rev.\ D {\bf 11} (1975)
2856

\bibitem{Kondo:2000sh}
K.~I.~Kondo, hep-th/0009152

\bibitem{Kondo:2000zv}
K.~I.~Kondo and T.~Shinohara, Prog.\ Theor.\ Phys.\  {\bf 105}
(2001) 649

\bibitem{Kondo:2000ey}
K.~I.~Kondo and T.~Shinohara, Phys.\ Lett.\ B {\bf 491} (2000) 263

\bibitem{Kondo:1999tj}
K.~I.~Kondo and Y.~Taira, Prog.\ Theor.\ Phys.\  {\bf 104} (2000)
1189

\bibitem{Kondo:1999je}
K.~I.~Kondo, Int.\ J.\ Mod.\ Phys.\ A {\bf 16} (2001) 1303

\bibitem{Kondo:1997kn}
K.~I.~Kondo, Prog.\ Theor.\ Phys.\ Suppl.\  {\bf 131} (1998) 243

\bibitem{Kondo:1998sr}
K.~I.~Kondo, Phys.\ Rev.\ D {\bf 58} (1998) 105019

\bibitem{Kondo:1997pc}
K.~I.~Kondo, Phys.\ Rev.\ D {\bf 57} (1998) 7467

\bibitem{Freire:2002ba}
F.~Freire, hep-th/0205032

\bibitem{Freire:2001nd}
F.~Freire, Phys.\ Lett.\ B {\bf 526} (2002) 405

\bibitem{Schaden:2001xu}
M.~Schaden, ''Wien 2000, Quark confinement and the hadron spectrum
258-268'', hep-th/0108034

\bibitem{Schaden:2000fv}
M.~Schaden, hep-th/0003030

\bibitem{Schaden:1999ew}
M.~Schaden, hep-th/9909011

\bibitem{Dudal:2002aj}
D.~Dudal, K.~Van Acoleyen and H.~Verschelde, ''Confinement,
Topology and Other Non-Perturbative Aspects of QCD, NATO Science
Series'' (2002), 97-104, hep-th/0204216

\bibitem{Min:bx}
H.~Min, T.~Lee and P.~Y.~Pac, Phys.\ Rev.\ D {\bf 32} (1985) 440

\bibitem{Coleman:jx}
S.~R.~Coleman and E.~Weinberg, Phys.\ Rev.\ D {\bf 7} (1973) 1888

\bibitem{Gross:jv}
D.~J.~Gross and A.~Neveu,
Phys.\ Rev.\ D {\bf 10} (1974) 3235.

\bibitem{Luperini:1991sv}
C.~Luperini and P.~Rossi,
Annals Phys.\  {\bf 212} (1991) 371.

\bibitem{Verschelde:jx}
H.~Verschelde, S.~Schelstraete and M.~Vanderkelen,
Z.\ Phys.\ C {\bf 76} (1997) 161.

\bibitem{Boucaud:2000}
P.~Boucaud, G.~Burgio, F.~Di Renzo, J.~P.~Leroy, J.~Micheli,
C.~Parrinello, O.~P$\grave{\textrm{e}}$ne, C.~Pittori,
J.~Rodriguez-Quintero, C.~Roiesnel and K.~Sharkey, JHEP {\bf 004}
(2000) 006

\bibitem{Boucaud:2001st}
P.~Boucaud, A.~Le Yaouanc, J.~P.~Leroy, J.~Micheli,
O.~P$\grave{\textrm{e}}$ne and J.~Rodriguez-Quintero, Phys.\ Rev.\
D {\bf 63} (2001) 114003

\bibitem{Gubarev:2000eu}
F.~V.~Gubarev, L.~Stodolsky and V.~I.~Zakharov, Phys.\ Rev.\
Lett.\ {\bf 86} (2001) 2220

\bibitem{Gubarev:2000nz}
F.~V.~Gubarev and V.~I.~Zakharov, Phys.\ Lett.\ B {\bf 501} (2001)
28

\bibitem{Verschelde:2001ia}
H.~Verschelde, K.~Knecht, K.~Van Acoleyen and M.~Vanderkelen,
Phys.\ Lett.\ B {\bf 516} (2001) 307

\bibitem{Boucaud:2002nc}
P.~Boucaud, J.~P.~Leroy, A.~Le Yaouanc, J.~Micheli,
O.~P$\grave{\textrm{e}}$ne, F.~De Soto, A.~Donini, H.~Moutarde and
J.~Rodriguez-Quintero, Phys.\ Rev.\ D {\bf 66} (2002) 034504

\bibitem{Kondo:2001nq}
K.~I.~Kondo, Phys.\ Lett.\ B {\bf 514} (2001) 335

\bibitem{Fukuda:1977mz}
R.~Fukuda and T.~Kugo, Prog.\ Theor.\ Phys.\  {\bf 60} (1978) 565

\bibitem{Fukuda:1977zp}
R.~Fukuda, Phys.\ Lett.\ B {\bf 73} (1978) 33 Erratum-ibid.\ B
{\bf 74} (1978) 433

\bibitem{Kondo:2001tm}
K.~I.~Kondo, T.~Murakami, T.~Shinohara and T.~Imai, Phys.\ Rev.\ D
{\bf 65} (2002) 085034

\bibitem{Dudal:2002ye}
D.~Dudal, V.~E.~Lemes, M.~S.~Sarandy, S.~P.~Sorella and
M.~Picariello,
JHEP {\bf 0212} (2002) 008.

\bibitem{Sawayanagi:zw}
H.~Sawayanagi,
Prog.\ Theor.\ Phys.\  {\bf 106} (2001) 971.

\bibitem{no}  K. Nakanishi and I. Ojima, \emph{Z. Phys. }\textbf{C6}(1980)
155

\bibitem{ni}  K. Nishijima, \emph{Prog. Theor. Phys. }\textbf{72 }(1984)
1214; \textbf{73} (1985) 536.

\bibitem{dj}  R. Delbourgo and P.D. Jarvis, \emph{J. Phys. }\textbf{A15}
(1982) 611; L. Baulieu and J. Thierry-Mieg, \emph{Nucl. Phys.
}\textbf{B197 }(1982) 477.

\bibitem{oji}  I. Ojima, \emph{Z.Phys. }\textbf{\ C13} (1982) 173.

\bibitem{ds}  F. Delduc and S.P.\ Sorella, \emph{Phys. Lett. }\textbf{B231}
(1989) 408.

\bibitem{cf}  G. Curci and R. Ferrari, \emph{Nuovo Cim. }\textbf{A32 }(1976)
151; \emph{Phys. Lett. }\textbf{B63} (1976) 91.

\bibitem{book}  O. Piguet and S.P. Sorella, \emph{Algebraic Renormalization}%
, Monograph series \textbf{m28}, Springer Verlag, 1995.\emph{\ }

\bibitem{bbh}  G. Barnich, F. Brandt and M. Henneaux, \emph{Phys. Rept.}
\textbf{338 }(2000) 439.

\bibitem{Lemes:2002ey}
V.~E.~Lemes, M.~S.~Sarandy and S.~P.~Sorella, hep-th/0206251


\bibitem{bo}  A. W. Overhauser, \emph{Advances in Physics }\textbf{27 }%
(1978) 343.

\bibitem{bezig}
D.~Dudal, H.~Verschelde,V.~E.~Lemes, M.~S.~Sarandy, S.~P.~Sorella
and M.~Picariello, paper in preparation.

\bibitem{Lemes:2002rc}
V.~E.~Lemes, M.~S.~Sarandy and S.~P.~Sorella,
hep-th/0210077.

\bibitem{Kondo:2002xn}
K.~I.~Kondo and T.~Imai, hep-th/0206173

\bibitem{Gracey:2001}
J.~A.~Gracey, Phys.\ Lett.\ B {\bf 525} (2002) 89

\bibitem{Browne:2002wd}
R.~E.~Browne and J.~A.~Gracey, Phys.\ Lett.\ B {\bf 540} (2002) 68

\bibitem{Gracey:2002}
J.~A.~Gracey, Phys.\ Lett.\ B {\bf 552} (2003) 101

\end{thebibliography}
\end{document}